\documentclass[%
 reprint,
 amsmath,amssymb,
 aps,
]{revtex4-2}

\usepackage{graphicx}% Include figure files
\usepackage{dcolumn}% Align table columns on decimal point
\usepackage{bm}% bold math

\begin{document}

\preprint{APS/123-QED}

\title{Mapping the phase diagram of a YBa$_2$Cu$_3$O$_{7-\delta}$ nanowire through electromigration}

\author{Edoardo Trabaldo}
\author{Alexei Kalaboukhov}
\author{Riccardo Arpaia}
\author{Eric Wahlberg}
\author{Floriana Lombardi}
\author{Thilo Bauch}
\email{bauch@chalmers.se}
\affiliation{Quantum Device Physics Laboratory, Department of Microtechnology and Nanoscience, Chalmers University of Technology, SE-41296 G\"{o}teborg, Sweden}\looseness=-1

\date{\today}% It is always \today, today,

\begin{abstract}

We use electromigration (EM) to tune the oxygen content of YBa$_2$Cu$_3$O$_{7-\delta}$ (YBCO)  nanowi\-res. During EM, the dopant oxygen atoms in the nanowire are moved under the combined effect of electrostatic force and Joule heating. The EM current can be tuned to either deplete or replenish nanowires with oxygen, allowing fine tuning of its hole doping level. Electrical transport measurements and Kelvin probe microscopy corroborate good homogeneity of the doping level  along the electromigrated nanowires. Thus, EM provides an effective method to study transport properties of YBCO in a wide doping range at the nanoscale in one and the same device.

\end{abstract}


\maketitle

%\tableofcontents

\section{Introduction}

The discovery of copper oxide-based High-critical Temperature Superconductors (HTSs), more than 30 years ago, ranks among the major scientific events in solid state physics and has given a tremendous boost to the field of superconductivity. The strong electron-electron correlations of these compounds lead to a plethora of symmetry breaking orders, such as charge and spin density waves, which strongly depend on the doping level of these materials \cite{keimer2015quantum}. The rather similar energy scales of those orders result in an intertwined ground state where charge, spin, lattice, and orbital degrees of freedom are entangled in a complex manner leading to an intricate temperature-doping phase diagram. The least understood phase, the so-called strange metal phase, is characterized by a linear-in-$T$ resistivity at temperatures above the pseudogap temperature. This phase is suggested to originate from a putative quantum critical point close to optimal doping \cite{ramshaw2015quasiparticle, varma2020colloquium}, which possibly promotes superconductivity in these materials. The study of HTSs in the form of nanowires \cite{lombardi1998transport, MOHANTY2004666,bonetti2004electronic, carillo2010little} and nanodots \cite{gustafsson2013fully} at various doping levels is expected to shine light on the microscopic mechanism leading to superconductivity in these materials, which emerges from such a complex ground state below the superconducting transition temperature $T_{\mathrm{c}}$. Here, investigating transport properties on length scales comparable to correlation length scales of various symmetry breaking orders  will be instrumental to understand the interplay or intertwining between those orders and superconductivity \cite{fradkin2015colloquium, auvray2019nematic, loret2019intimate, wahlberg2021restored, huang2021two}. At the same time nanoscale HTSs allow for the development of quantum limited sensors, such as nanoscale superconducting quantum interference devices \cite{arpaia2014ultra, schwarz2015low, charpentier2016hot, arzeo2016toward, martinez2018nanosquid,  trabaldo2019grooved, li2020high, trabaldo2020properties} and  photon detectors \cite{curtz2010patterning, arpaia2017transport, shibata2017photoresponse, ejrnaes2017observation, amari2017high, andersson2020fabrication}, whose properties can be strongly modified by changing the doping level. 

Typically, nanowires with different doping levels are obtained by nanopatterning of thin films  having the corresponding doping level, i.e. oxygen content, in the case of YBa$_2$Cu$_3$O$_{7-\delta}$ (YBCO) \cite{andersson2020fabrication,arpaia2017transport}. However, such a fabrication process is very challenging, especially for lower doping levels, and requires a separate YBCO film for each doping level \cite{andersson2020fabrication}. Moreover, this approach makes the comparison between transport properties of nanowires of different doping levels difficult since properties such as critical current sensitively depend on wire morphology on the nanoscale \cite{nawaz2013microwave}, which will unavoidably vary between devices.

\begin{figure}[b]
\includegraphics[width=0.48\textwidth]{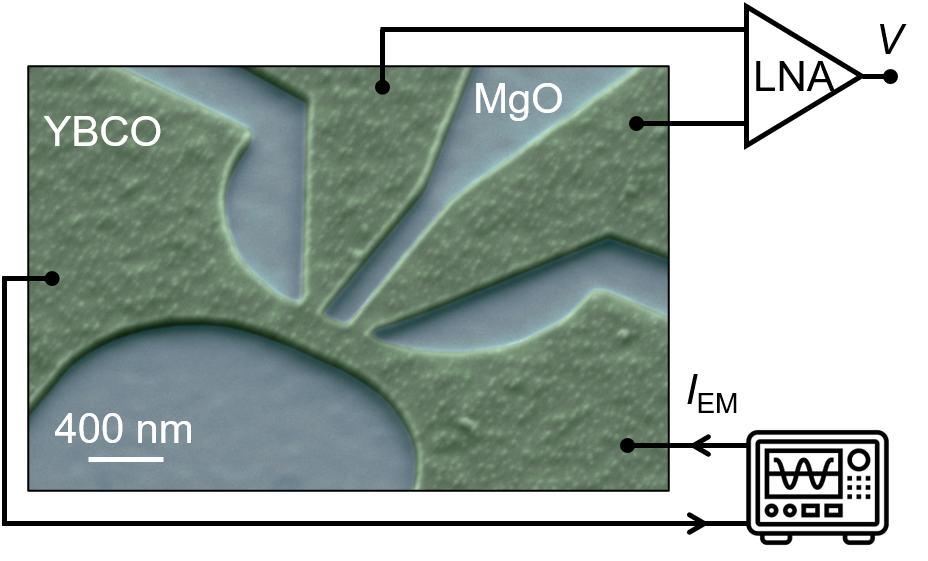}
\caption{Scanning electron microscope (SEM) image of a typical nanowire used for electromigration. An arbitrary waveform generator supplies the electromigration current $I_\mathrm{EM}$ and a low noise amplifier (LNA) is used to simultaneously probe the resistance of the nanowire.}
\label{Fig:1}
\end{figure}
Tuning the doping of YBCO ex-situ would solve the limitations inherent to fabrication. One way to tune the material's electrical properties is via electric field effect or gating. Although moderate success has been achieved, most gating techniques require either the use of ferroelectric materials \cite{lemanov1993ybco,crassous2013bifeo3} or electrolytes \cite{zhang2017mechanism}, in combination with strong electric fields. Moreover, the effectiveness of gating is tied to the thickness of the material and limits its application to ultrathin films \cite{xi1991electric}.

Another promising route to modify the doping level in one and the same structure is to use electromigration (EM) \cite{moeckly1993electromigration,baumans2019electromigration}. Here the dopant oxygen atoms are moved by subjecting the material to large current densities in the normal state, thus tuning the doping level ex-situ. YBCO is a particularly good candidate for EM due to the low activation energy of the dopant oxygen atoms \cite{scouten1994low}. However, electromigration on $\mu$m wide structures still suffer from rather inhomogeneous doping distributions along YBCO bridges \cite{Marinkovic2020}. 

In this work we use EM to tune the doping level of YBCO nanowires. We find that the doping can be increased or decreased depending on the waveform of the applied bias current. By applying a DC current we can replenish the oxygen in the nanostructures, while an AC current can be used to deplete it. The effect of the EM is studied by measuring the resistance versus temperature of the nanowires before and after the process. The resulting curves generally show a single sharp superconductive transition, which indicates homogeneity of the doping distribution. Successive steps of EM on a single nanowire are used to recreate most of the phase diagram of YBCO, covering a wide range of doping. Finally, Kelvin Probe Force Microscopy (KPFM) preformed on nanowires after EM confirms the homogeneous doping distribution along the wire.

\section{Results}

The quality of the pristine YBCO nanowires is of paramount importance since defects are affected by the electromigration process and may worsen the quality of the nanowire with each EM step. The YBCO nanowires used in this work have been fabricated from thin films (close to the optimal doping) with thickness of $50$~nm. The thin films were first deposited by pulsed laser deposition (PLD) onto (110) oriented MgO substrates, and then covered by a hard layer of carbon, also deposited by PLD. Electron beam lithography and oxygen plasma etching are used to define the carbon mask, which is then used to etch the thin films in the final shape of the nanowires through Ar ion milling. Finally, the carbon mask is removed by oxygen plasma etching. This fabrication has been shown to yield high quality nanowires, both for optimally- \cite{nawaz2013approaching, arpaia2016improved, trabaldo2017noise} and under-doped \cite{andersson2020fabrication} films.

A scanning electron microscope (SEM) image of a nanowire is shown in Fig.~\ref{Fig:1}. The nanowires are patterned with a 4-point geometry, with $200$~nm width and $400$~nm length (distance between voltage probes). We use rounded inner corners at the end points of the nanowires to mitigate the effect of current crowding \cite{hagedorn1963right,clem2011geometry,nawaz2013microwave}, which could locally increase the effect of the EM process. In Fig.~\ref{Fig:1} is also reported the schematic of the electromigration and measurement setup connected to the nanowire. An arbitrary waveform generator (AWG) is connected to the nanowire and supplies the EM current, while the voltage drop along the nanowire is amplified by a low noise amplifier (LNA).

The electromigration is typically performed at room temperature, although it can be done also at lower temperature, $T$, without significant changes. We perform two types of electromigration: DC-EM, where the AWG is maintained at a DC bias value during the whole process, and AC-EM, where the AWG output is an ambipolar square wave signal, i.e. a pulse train. These two types of EM waveforms have different effects on the nanowire doping as will be discussed below.

\begin{figure}[t]
\includegraphics[width=0.48\textwidth]{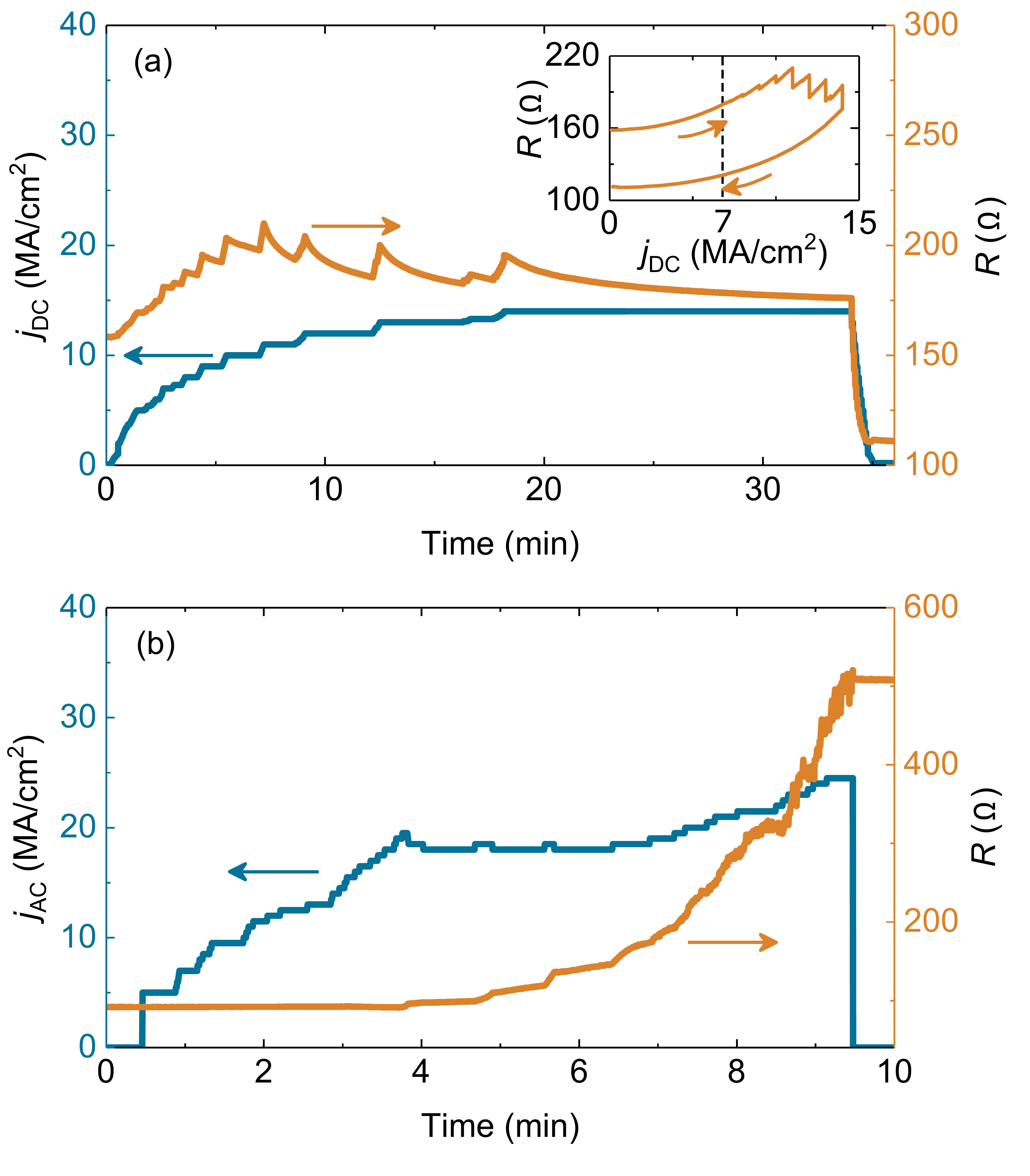}
\caption{(a) DC-EM process. DC current density, $j_\mathrm{DC}$ (left axis), and resistance, $R$ (right axis) as a function of time during DC-EM. In the inset, the resistance is shown as a function of the bias current used in the DC-EM process. (b) AC-EM process. AC current density , $j_\mathrm{AC}$ (left axis), and resistance, $R$ (right axis) as a function of time during AC-EM. }
\label{Fig:1p5}
\end{figure}

An example of DC-EM is shown in Fig.~\ref{Fig:1p5}(a). We start with a nearly optimally doped nanowire with hole doping $p\simeq 0.16$ and resistance $R=159$~$\Omega$. Here the doping level is estimated by comparing the superconducting transition temperature $T_{\mathrm{c}}$ of the nanowire to that of YBCO films of known hole doping \cite{arpaia2018probing}. The DC bias current density across the wire, $j_\mathrm{DC}$, is manually increased while $R$ is monitored. For low values of $j_\mathrm{DC}$ ($<7$~MA/cm$^2$) the resistance increases approximately quadratically with the current density, see inset of Fig. \ref{Fig:1p5} (a). This behavior is caused by the local temperature increase in the wire $\Delta T \propto Rj_\mathrm{DC}^2$, due to Joule heating, and the linear in $T$ resistance of optimally doped YBCO.

The bias current is further increased until it is sufficiently high to start the EM of the dopant oxygen atoms, which happens for $j_\mathrm{DC}\simeq 7$~MA/cm$^2$. In this regime, with  $j_\mathrm{DC}$ kept constant, the nanowire resistance slowly decreases over time. A reduction in $R$ indicates that the doping level of the nanowire is increased \cite{arpaia2018probing} and oxygen is being replenished in the nanostructure. Once $j_\mathrm{DC}$ is increased even further, $R$ starts decreasing faster (see Fig.~\ref{Fig:1p5}(a)). After the DC-EM, i.e. when $j_\mathrm{DC}$ is brought back to $0$, the nanowire resistance drops to $R=111$~$\Omega$. From this value and the superconducting transition temperature (data not shown) we estimate that the doping level inside the wire has increased to a value $p\simeq 0.175$ \cite{arpaia2018probing}.

However, the bias current used for DC-EM cannot be increased indefinitely. We observe that if $j_\mathrm{DC}$ is raised above a certain threshold ($>15$~MA/cm$^2$), the DC-EM causes a depletion of the oxygen atoms within the wire. This means that at high bias, DC-EM lowers the nanowire doping instead of increasing it. Moreover, when the oxygen is reduced by DC-EM, the resulting nanowires show inhomogeneous transport properties, e.g. multiple superconducting transitions (data not shown). Therefore, in this work DC-EM is only used for increasing the doping of YBCO nanowires.

To achieve homogeneous reduction of the doping level one can instead use AC-EM, which results in a controlled depletion of oxygen in the nanowires. The bias current waveform used for AC-EM is a square wavefunction centered around zero, with peak to peak current density $j_\mathrm{AC}$ and frequency $f_\mathrm{AC}=1$ kHz. At this frequency the effect of electromigration is negligible due to the relatively long oxygen diffusion relaxation times \cite{diff_note}. Instead, thermomigration of oxygen atoms due to large temperature gradients between the wire and the electrodes dominates the oxygen migration process in YBCO nanowires \cite{oriani1969thermomigration}. Here, similar to the DC-EM, the resistance of the nanowire is measured  between 100~ms long AC-EM bursts to monitor the effect of the electromigration. An example of AC-EM is shown in Fig.~\ref{Fig:1p5}(b). Here we start with a nanowire with $p\simeq 0.175$ and $R=92$~$\Omega$. When $j_\mathrm{AC}$ is sufficiently high ($\simeq 17$~MA/cm$^2$), $R$ starts to increase. Increasing $j_\mathrm{AC}$ further causes a faster increase of the wire resistance $R$. 
During AC-EM, $j_\mathrm{AC}$ needs to be increased slowly to avoid thermal runaway related to the abrupt increase of $R$, which would be detrimental for the nanowire. In Fig.~\ref{Fig:1p5}(b) the bias current is adjusted multiple times during the EM to avoid thermal runaway. After the AC-EM process, the nanowire resistance has increased to $R=508$~$\Omega$ corresponding to a reduction of the doping value to $p=0.08$ (determined from $T_{\mathrm{c}}$ of the resistive transition \cite{arpaia2018probing}). 

\begin{figure}[]
\includegraphics[width=0.48\textwidth]{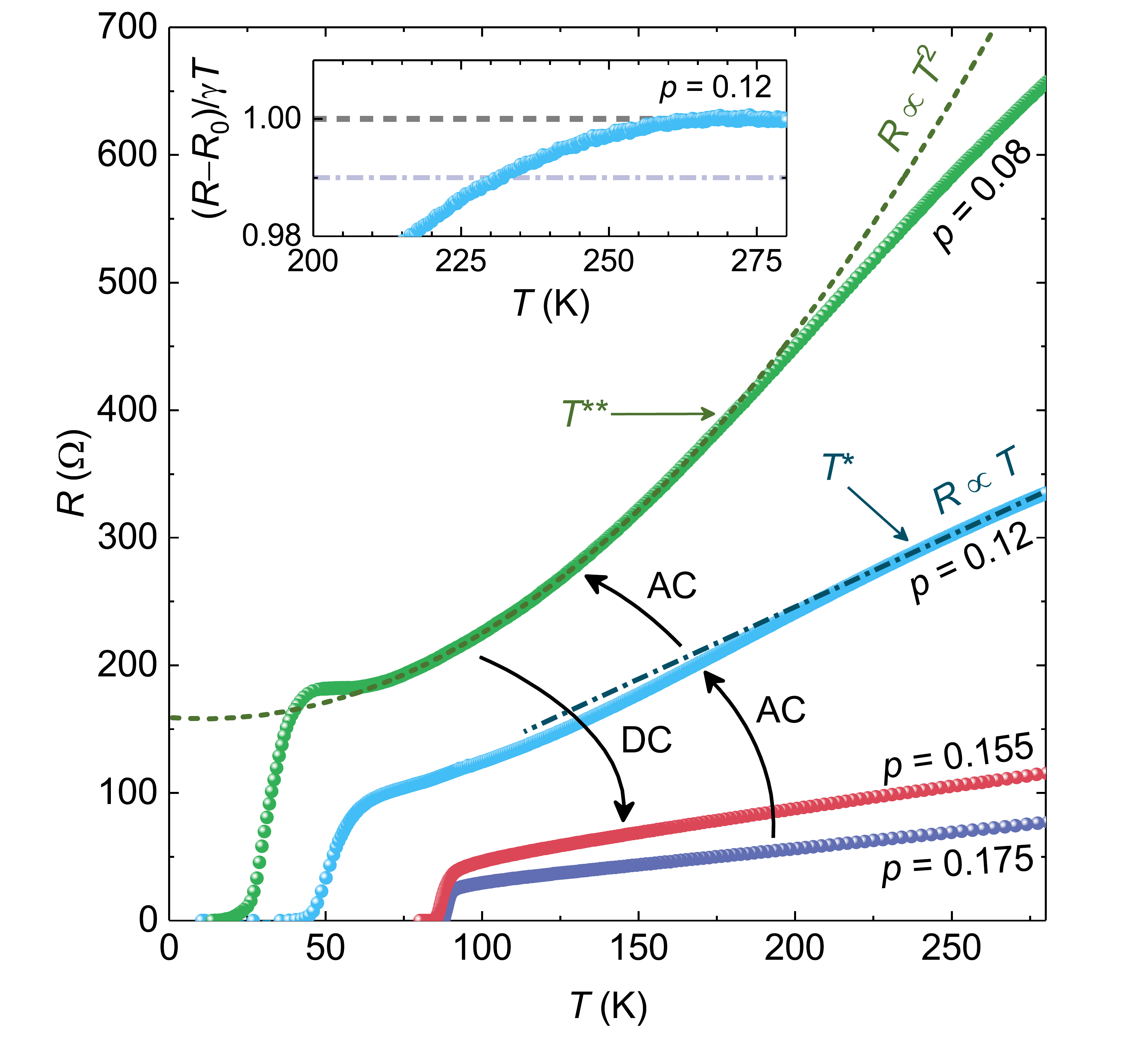}
\caption{Evolution of the $R$($T$) measured on the same nanowire after successive electromigration steps. The initial doping, $p=0.175$, is first decreased to $p=0.12$ and then to $p=0.08$ through AC-EM. The doping is then increased to $p=0.155$ using DC-EM. The dashed line is a quadratic fit of the $R(T)$ data at $p=0.08$. $T^{**}$ is the temperature above which the resistance departs from the quadratic in $T$ behavior. The dash-dotted line is a linear fit to the $R(T)$ data at $p=0.12$. $T^{*}$ is the temperature below which the resistance departs from the linear in $T$ behavior. Inset: $(R-R_0)/\gamma T$ as a function of temperature for $p=0.12$, where $\gamma=0.65~\Omega/\mathrm{K}$ and $R_\mathrm{0}=35.5~\Omega$ are the fitted coefficients of the linear in T fit at high temperatures $\gamma T + R_\mathrm{0}$. $T^{*}$ is given by the temperature below which the measured $R$ departs by more than $1\%$ (see dash dotted line) from the liner in T dependence.% difference between the measured $R(T)$ at $p=0.12$ and the linear fit $\gamma T + R_\mathrm{0}$, with $\gamma=0.65$ and $R_\mathrm{0}=35.5$; the lighter dot-and-dashed line is placed at $1\%$ deviation as a guide for the eye.}
}
\label{Fig:2}
\end{figure}

In Fig.~\ref{Fig:2} we show four measurements of resistance versus temperature, $R(T)$, done on the same nanowire to compare the effect of DC- and AC-EM. Starting from a slightly overdoped nanowire, $p=0.175$, we apply first AC-EM  to lower the doping to $p=0.12$, and a second time to further lower the nanowire doping to $p=0.08$. The decrease in doping results in an increase of the resistance, and a lower critical temperature $T_\mathrm{c}\simeq32$~K. DC-EM is then used to increase the nanowire doping up to $p\simeq0.155$, i.e. in the optimally doped regime of YBCO. Higher $p$ is accompanied by higher $T_\mathrm{c}\simeq87$~K and lower value of resistance. It is important to note that the $R(T)$ curves after EM (both AC and DC) present a single superconductive transition, which indicates homogeneity of the doping along the nanowire. Thus, a combination of DC- and AC-EM can be used to increase or to decrease the doping of our nanowires, respectively, and therefore fine tune the doping level.
\begin{table}[]
    \centering
    \begin{tabular}{r|c|c|c|c}
         & doping $p$ & $T^*$ [K] & $T^{**}$ [K] & $\Delta T_\mathrm{c}$ [K]\\
        \hline
        \hline
        nanowire & 0.155 & 148 & - & 4.5\\
        film & 0.155 & 145 & - & 3.4\\
        \hline
        nanowire & 0.12 & 230 & 120 & 14\\
        film & 0.12 & 215 & 118 & 6\\
        \hline
        nanowire & 0.08 & 271 & 170 & 14\\
        film & 0.08 & 263 & 170 & 7
    \end{tabular}
    \caption{Characteristic temperatures $T^*$, $T^{**}$, and $\Delta T_\mathrm{c}$ (see text for details) extracted from a single electromigrated nanowire (see Fig.~\ref{Fig:2}) compared to the values reported for YBCO thin films \cite{arpaia2018probing}.}
    \label{tab:1}
\end{table}

To confirm the homogeneity of the nanowires we can first extract the superconducting transition width, $\Delta T_\mathrm{c}$, which is defined as the temperature range in which the wire resistance changes from $90\%$ to $10\%$ of the normal state resistance \cite{arpaia2018probing}. The values of $\Delta T_\mathrm{c}$ extracted for our nanowires are reported in Table~\ref{tab:1}, with the values obtained in thin films at similar doping. Only when we reach the strongly underdoped regime, the superconducting transitions of the nanowires become wider in $T$ than for the thin films.

Of particular interest are the various characteristic temperatures, which can be extracted from the $R(T)$ curve of a YBCO nanowire. These temperatures can be compared to those extracted from YBCO thin films \cite{arpaia2018probing} to further asses the quality of our nanowires after electromigration. Here we focus on two characteristic temperatures, the first is $T^*$, i.e. the temperature below which the $R(T)$ departs from linearity ($R\propto T$), as indicated in Fig.~\ref{Fig:2} for $p=0.12$. The second characteristic temperature is $T^{**}$, i.e. the upper bound of the $T$ range in which $R\propto T^2$, as shown in Fig.~\ref{Fig:2} for $p=0.08$. In the inset of Fig.~\ref{Fig:2} we show the relative deviation of the measured $R(T)$ at $p=0.12$ from the linear in $T$ dependence. The value of $T^*$ is determined as the temperature below which the $R(T)$ departs by more than $1\%$ from the linear-in-$T$ dependence, indicated in the inset by the dash-dotted line.

In Table~\ref{tab:1} the values of $T^*$, $T^{**}$ and $\Delta T_\mathrm{c}$ extracted from the $R(T)$ curves from Fig.~\ref{Fig:2} are compared to the values reported for YBCO thin films at similar doping \cite{arpaia2018probing}. Indeed, from this table one can see only small variations between the thin-film and our nanowire properties. After the electromigration, the nanowire properties are therefore still close to pristine YBCO films. Such occurrence  highlights that the effects of the electromigration are limited to the displacement of oxygen atoms, without degradation of the underlying material, resulting in homogeneous YBCO nanowires in a broad doping range.

\begin{figure}[t]
\includegraphics[width=0.48\textwidth]{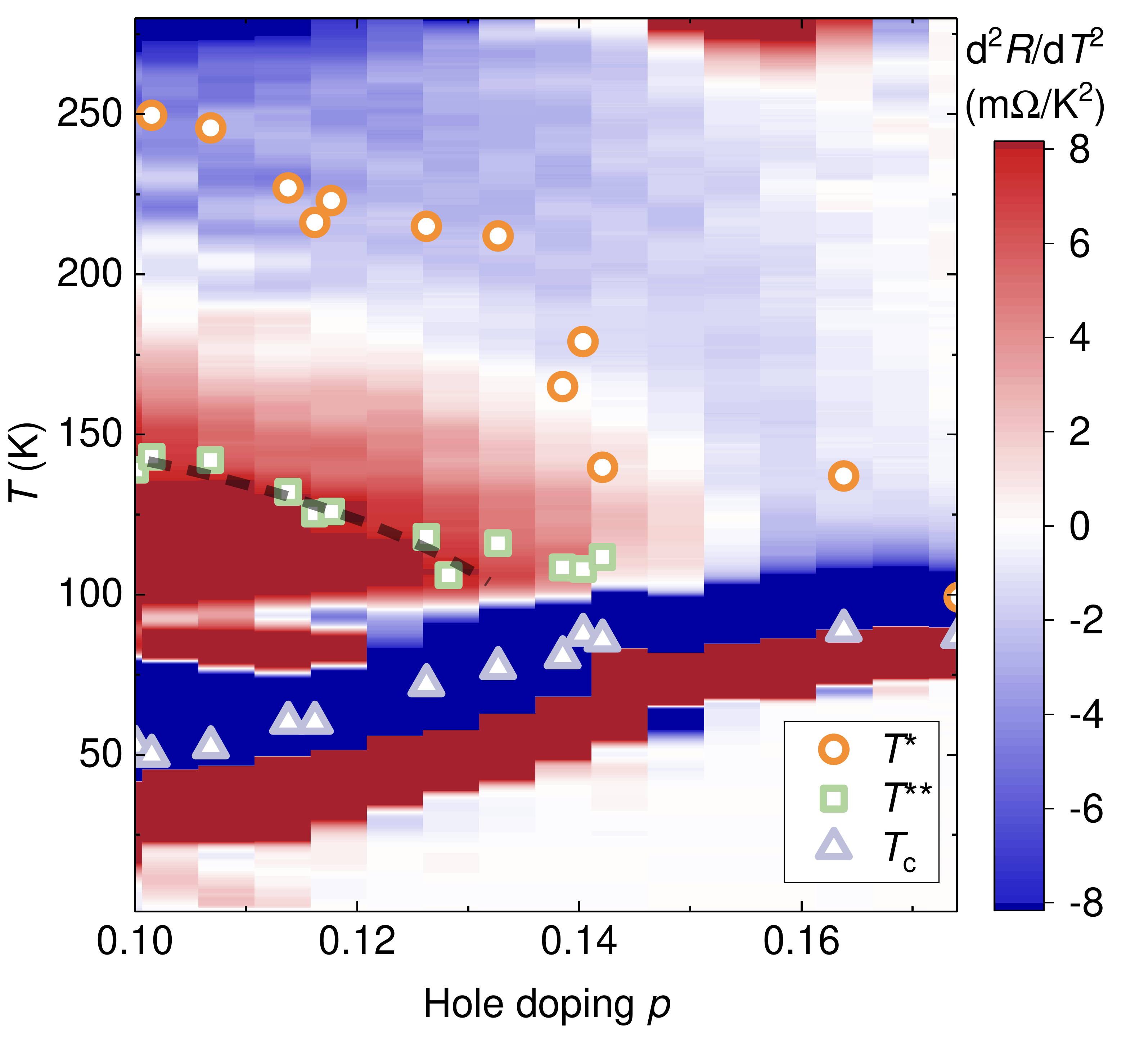}
\caption{Second derivative of the resistance, $d^2R/dT^2$, versus temperature and doping, obtained on the same nanowire after successive electromigration steps. The values of $T_\mathrm{c}, T^*$ and $T^{**}$ measured on thin films of various doping levels have been adapted from Ref. \citenum{arpaia2018probing}. Note, the feature (horizontal white line) around 90~K for doping levels below $p=0.13$ is caused by the proximity of the superconducting transition in the voltage probes, which are connected to the wire (see Fig.~\ref{Fig:1}). The voltage probes are much less affected by the EM process and therefore retain a $T_{\mathrm{C}}$ value close to the pristine wire. The dashed line is placed at the $T^{**}$ values of the electromigrated nanowires as a guide for the eye.}
\label{Fig:3}
\end{figure}
By performing AC-EM on a single nanowire we reproduced the phase diagram of YBCO shown in Fig.~\ref{Fig:3}. The nanowire was fabricated with with $p=0.175$. Its doping was reduced in successive steps of AC-EM, while the $R(T)$ was measured after each step. In Fig.~\ref{Fig:3} we show the second derivative of the measured resistance versus temperature, $d^2R/dT^2$, as a function of temperature and doping. This phase diagram is in good agreement with the one obtained from YBCO thin films \cite{arpaia2018probing} and bulk samples \cite{ando2004electronic,barivsic2013universal}. One can clearly identify  the linear in $T$ resistivity behavior in the strange metal phase slightly above the optimal doping level (see white colored region for $p>0.16$ and $T>T_{\mathrm{C}}$ in Fig.~\ref{Fig:3}).

Moreover, for comparison, the values of $T^*$, $T^{**}$ and $T_\mathrm{c}$ extracted from measurements performed on YBCO thin films \cite{arpaia2018probing} have been added to Fig.~\ref{Fig:3} (open symbols). The values of $T^*$ extracted from the $R(T)$ curves of our nanowires lie close to the thin films values (orange circles). However, the phase diagram shows that the white region, which corresponds to purely linear $R(T)$, is well above these values. A similar behavior has been reported for bulk crystals \cite{ando2004electronic}. Overall, we find a good agreement between our phase diagram and the one reported in  literature for bulk crystals \cite{ando2004electronic}, confirming that EM can be used for tuning ex-situ the doping of individual YBCO nanowires while preserving electrical properties close to those of pristine thin films and single crystals.

\begin{figure}[t]
\includegraphics[width=0.48\textwidth]{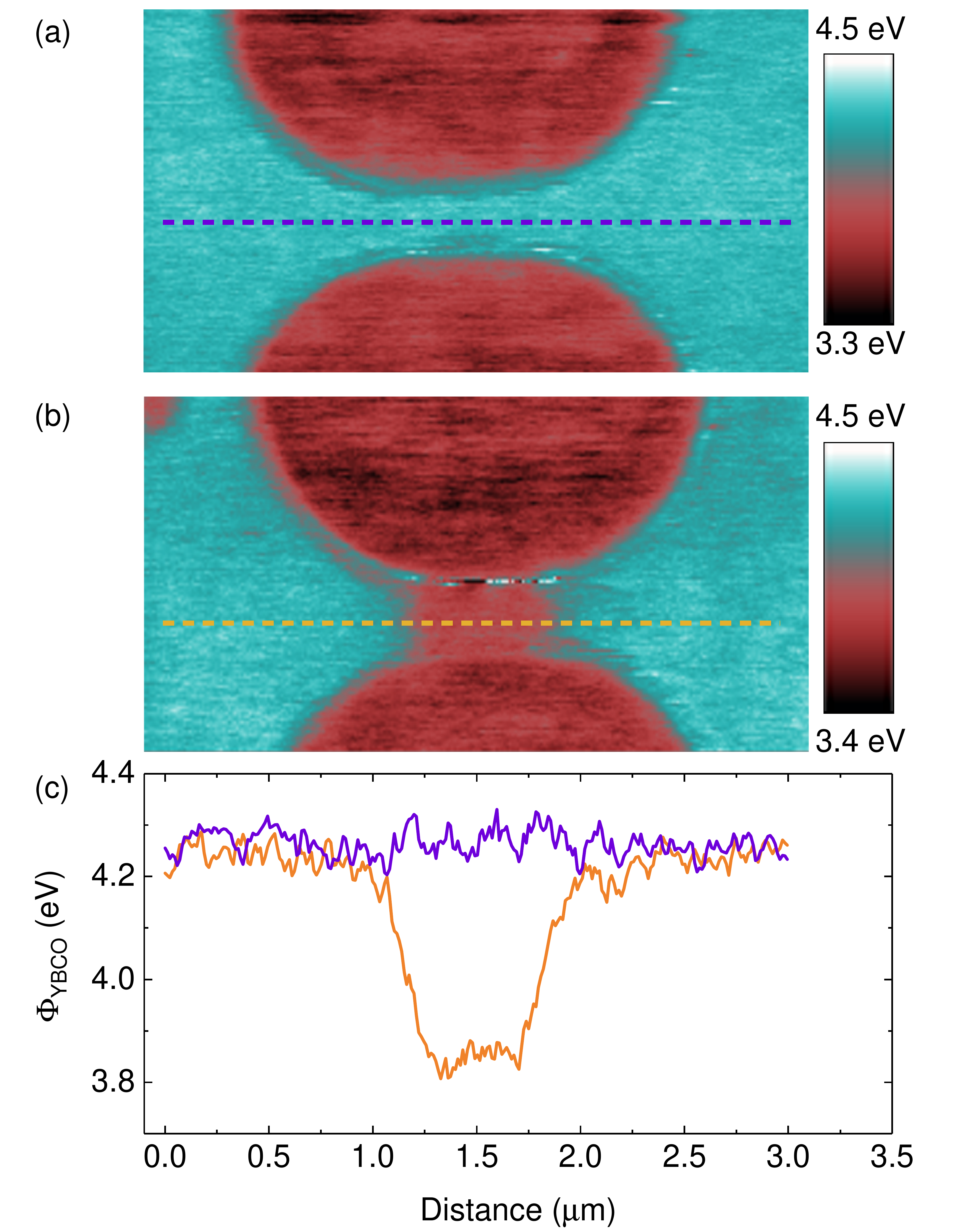}
\caption{Work function map of a YBCO nanowire extracted from KPFM images before EM (slightly overdoped) (a) and after EM (underdoped) (b). (c) Line cuts of the work function along the YBCO wire for the optimally doped wire (blue line, see also dashed blue line in panel (a)) and underdoped nanowire after AC-EM (red line,  see also dashed red line in panel (b)).}
\label{Fig:4}
\end{figure}

To further investigate the local effect of EM on the nanowires, we employ Kelvin probe force microscopy (KPFM) to visualize the doping distribution after EM with nanometer resolution. KPFM is a variant of non-contact atomic force microscopy (AFM), which measures the difference in work function, $\Phi$, between the nanowire surface and the conductive AFM tip \cite{melitz2011kelvin}.

Two nanowires, one slightly overdoped (as-fabricated, $p=0.175$) and one underdoped ($p\simeq 0.05$, obtained through AC-EM),  have been imaged. In order to obtain absolute values of the work function in YBCO,  $\Phi_{\mathrm{YBCO}}$, we have performed a calibration of the AFM tip work function, $\Phi_{\mathrm{tip}}$, using a highly oriented pyrolytic graphite sample (HOPG): $\Phi_{\mathrm{tip}} = eV_{\mathrm{CPD}}+\Phi_{\mathrm{HOPG}}$, where $V_{\mathrm{CPD}}$ is the measured potential difference between the AFM tip and the HOPG surface, and $\Phi_{\mathrm{HOPG}} = 4.48$~eV is the work function of HOPG. The work function of the YBCO nanowire can then be obtained as $\Phi_{\mathrm{YBCO}} = \Phi_{\mathrm{tip}}  - eV_{\mathrm{CPD}}$, where $V_{\mathrm{CPD}}$ is the measured surface potential difference between the tip and the YBCO nanowire. The measured KPFM maps of $\Phi_{\mathrm{YBCO}}$ of the YBCO nanowire before and after EM treatment are shown in Fig. \ref{Fig:4}. The absolute value of $\Phi_{\mathrm{YBCO}}\simeq 4.3$~eV is close to values reported earlier for optimally doped YBCO bulk polycrystalline \cite{Koshida_1990} and thin film \cite{Rietveld_1992} samples. In the EM-treated underdoped YBCO nanowire, the central region has significantly lower work function as compared with the wide electrodes. The value of the work function taken along the nanowire, i.e. along the linecuts in panel (a) and (b), are shown in panel (c). The difference in work function is about $0.5$~V between the slightly overdoped and underdoped YBCO nanowires. 

The work function is defined as a difference between the Fermi level of a metal and vacuum level just outside the surface. Thus the work function of YBCO, $\Phi_{\mathrm{YBCO}}$, and cuprates in general, is expected to be influenced by the doping, and by measuring the work function one can indirectly map the doping distribution in our nanowires. Indeed, the difference between work functions of optimally doped and underdoped YBCO nanowires is in good agreement with the chemical potential shift as a function of doping measured by X-ray photoemission spectroscopy \cite{maiti2009doping,yagi2010characteristic}. On the other hand, KPFM is a surface sensitive technique and the potential difference may also depend on a surface dipole layer due to the presence of adsorbates. A similar change in work function was found in YBCO films modified using scanning tunneling microscopy \cite{Urazhdin_2003}. The result was interpreted as due to the presence of trapped charges at the oxygen vacancies at the modified surface. While we cannot rule out that similar effects contribute to the difference in work function in our nanowires, we emphasize that the transport properties indicate that the modification occurs in the bulk of the YBCO rather than at the surface. 

\section{Conclusions}

We have presented an electromigration technique which can be used to fine tune the oxygen content, and thus the doping level, of YBCO nanowires. Depending on the bias current waveform used for the EM we can increase (DC-EM) or decrease (AC-EM) the nanowire doping. We used EM to reproduce most of the phase diagram of YBCO with a single nanowire. We used KPFM imaging to asses the effect of the EM on the morphology and work function of the nanowires after EM. This confirmed that the doping distribution in electromigrated nanowires is homogeneous over a consistent area. The ex-situ control of the doping level in YBCO nanowires is a milestone towards an in-depth study of the phase diagram of cuprates on the nanoscale and opens the way to investigate transport properties in the proximity of putative quantum critical points \cite{keimer2015quantum} in one and the same nanowire where fine tuning of the doping level is required. %The EM bypasses the challenges of different fabrication techniques and opens the way for using the same nanowire to study the entire doping range. 
Moreover, this technique is interesting for the technological applications of YBCO weak links, where the tuning of the nanowire properties, such as critical current and kinetic inductance, is required.

\begin{acknowledgments}

This work has been performed in part at Myfab Chalmers and is supported by the Knut and Alice Wallenberg Foundation (KAW) under the project ``NeuroSQUID" (2014.0102\_KAW) and the Swedish Research Council (2020-05184\_VR, 2018-04658\_VR, 2020-04945\_VR).

\end{acknowledgments}

\bibliography{bibliography}% Produces the bibliography via BibTeX.

\end{document}